\documentclass[aps,twocolumn,superscriptaddress,amsmath]{revtex4-1}

\pdfoutput = 1

\usepackage[pdftex]{graphicx,color}
\usepackage{soul}
\usepackage{amsfonts}
\usepackage{amssymb}
\usepackage{amsmath}
\usepackage{amsthm}

\usepackage{subfigure}
\usepackage{rotate}
\usepackage{bm}
\usepackage{dcolumn}

\usepackage[pdftex]{hyperref}

\let\oldmarginpar\marginpar
\renewcommand\marginpar[1]{\-\oldmarginpar[\raggedleft\tiny #1]%
{\raggedright\tiny #1}}

\renewcommand{\Re}{{\mathrm{Re}}\, }

\newcommand{\bra}[1]{\langle#1|}
\newcommand{\ket}[1]{|#1\rangle}

\newcommand{\set}[1]{\{ #1 \}}
					
\graphicspath{{figs/}}

\begin{document}


\title{Many-body mobility edge in a mean-field quantum spin glass}
	

\author{C. R. Laumann}
\email[Correspondence: ]{claumann@uw.edu}
\affiliation{Department of Physics, University of Washington, Seattle, WA 98195, USA}
\affiliation{Perimeter Institute for Theoretical Physics, Waterloo, Ontario N2L 2Y5, Canada}

\author{A. Pal}
\affiliation{Department of Physics, Harvard University, Cambridge, MA 02138, USA}

\author{A. Scardicchio}
\email[On leave from: ]{Abdus Salam ICTP, Strada Costiera 11, 34151 Trieste, Italy}
\affiliation{Physics Department, Princeton University, Princeton, NJ 08542, USA}
\affiliation{Physics Department, Columbia University, New York, NY 10027, USA}
\affiliation{ITS, Graduate Center, City University of New York, New York, NY 10016, USA}
\affiliation{INFN, Sezione di Trieste, Via Valerio 2, Trieste 34151, Italy}
\date{\today}


\begin{abstract}
The quantum random energy model provides a mean-field description of the equilibrium spin glass transition. 
We show that it further exhibits a many-body localization - delocalization (MBLD) transition when viewed as a closed quantum system. 
The mean-field structure of the model allows an analytically tractable description of the MBLD transition using the forward-scattering approximation and replica techniques. 
The predictions are in good agreement with the numerics. 
The MBLD lies at energy density significantly above the equilibrium spin glass transition, indicating that the closed system dynamics freezes well outside of the traditional glass phase. 
We also observe that the structure of the eigenstates at the MBLD critical point changes continuously with the energy density, raising the possibility of a family of critical theories for the MBLD transition.
\end{abstract}

\maketitle





\paragraph{Introduction---}
Equilibrium statistical mechanics applied to closed dynamical systems relies on the assumption of ergodicity. Until recently it was believed that even weak interaction between the elementary constituents of matter guarantees ergodicity. 
A notable counterexample was provided by the seminal work \cite{Basko:2006hh} where the authors showed that, for systems with quenched disorder, Anderson localization of noninteracting particles \cite{anderson1958absence} can persist in the presence of (sufficiently weak) interactions leading precisely to the failure of ergodicity. 
The recent development of well-isolated experimental quantum many-body systems has spurred a great deal of numerical and theoretical work suggesting that \emph{many-body localization} (MBL) is, indeed, a robust, universal phenomenon: it exists in any spatial dimension, for both bosons and fermions, and for generic short-range interactions \cite{oganesyan2007localization, vznidarivc2008many, pal2010MBL, aleiner2010finite, huse2013localization, pekker2013hilbert, vosk2013dynamical, bahri2013localization, chandran2013many, Aizenman:2009fp, serbyn:2013cl, Ioffe:2010dp,Vosk:2013kq, Kjall:2014fj, Imbrie:2014qv, Swingle:2013ys,Bauer:2013rw,Serbyn:2014eu,Yao:2013ly,Cuevas:2012ep}.
The MBL phase is characterized by the complete absence of transport (e.g.\ of particle, spin and energy) and by the permanence of the memory of the initial state in local observables for all time.
In this respect, the MBL phase may be viewed as the quintessential quantum glass. 

By changing the control parameters such as energy, strength of interactions or disorder, a MBL system can transit into a \emph{delocalized} phase, where transport is restored and the predictions of equilibrium statistical mechanics hold. 
In particular, the eigenstates satisfy the eigenstate thermalization hypothesis (ETH) \cite{Deutsch1991ETH, Srednicki1994ETH, rigol2008thermalization}, exhibiting thermal behavior for local observables. 
If the MBL-delocalization (MBLD) transition is obtained by changing just the energy of the state, all other parameters remaining the same, the critical energy density separating the two phases is also referred to, borrowing terminology from the theory of Anderson localization, as a many-body mobility edge.
When one passes from the micro canonical to the canonical statistical description, the mobility edge defines a critical temperature $T_{\text{MBL}}$. 


\begin{figure}[b]
  \centering
    \includegraphics{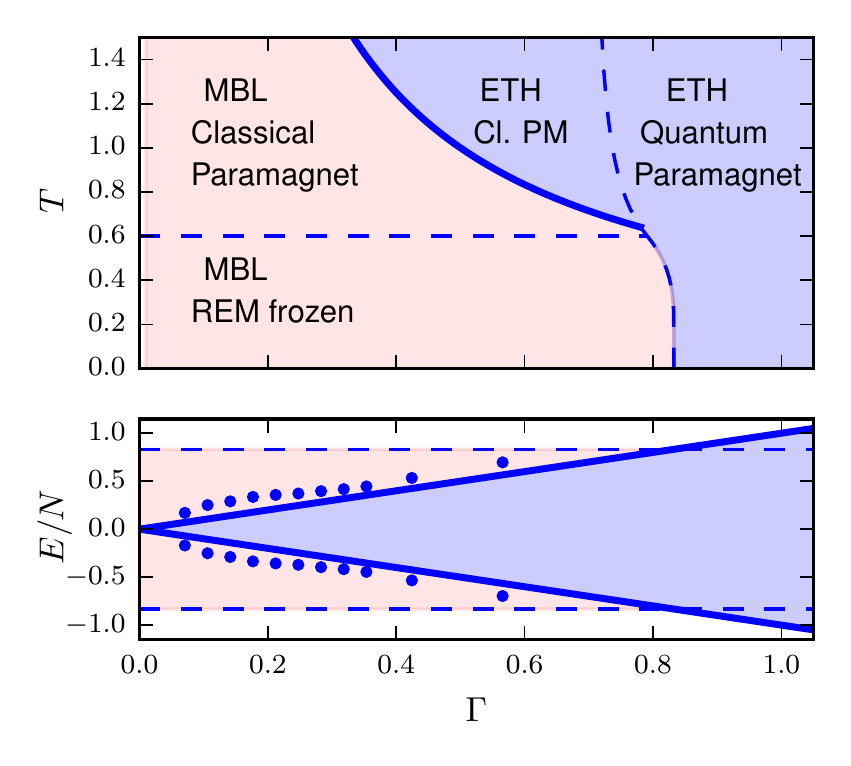}
  \caption{(a) The canonical phase diagram of the QREM in $\Gamma-T$ plane. Dashed lines correspond to first order thermodynamic transitions due to the crossing of free energies found in the replica treatment. Solid line corresponds to the MBLD transition at $T_{\text{MBL}} = 1/2\Gamma$. Red (blue) shaded region is localized (ergodic).
  (b) The microcanonical phase diagram in the $\Gamma-\epsilon$ plane. Shaded regions correspond to support of many-body spectrum. Blue dots are an estimate of the transition from finite-size crossing points of $[r]$ at fixed $\Gamma$.}
  \label{fig:pd}
\end{figure}

As MBL systems exhibit glassy dynamics, it is natural to ask whether the statistical models familiar from the theory of spin glasses \cite{Mezard:1987aa} actually exhibit MBLD transitions when endowed with quantum dynamics. 
And, in particular, if one of the various mean-field models of spin glass may provide an analytically tractable understanding of the MBLD transition. 
In this Letter, we show that this is true by studying the classical random energy model \cite{Derrida:1981jd} subjected to a transverse field $\Gamma$. 
Using a mixture of numerical and analytical techniques, we find a MBLD transition that does not coincide with the glass transition which was studied in previous works \cite{Goldschmidt:1990p962} (see also \cite{Jorg:2008fj}). 
Rather, the transition temperature lies strictly above the previously studied replica symmetry breaking `equilibrium' phase transition, dividing the (classical) paramagnetic phase into a paramagnet which is many-body localized and an ergodic one as seen in Figure \ref{fig:pd}.  

\begin{figure}[t]
  \centering
    \includegraphics{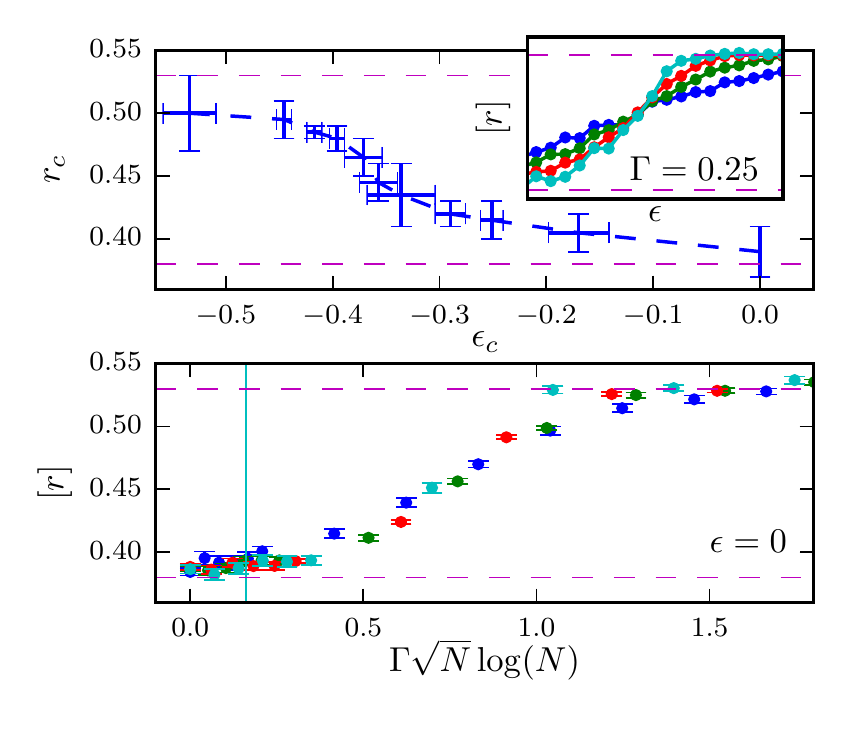}
  \caption{(color online) 
  (a) Critical value of level spacing ratio $r_c$ as a function of $\epsilon_c$ parametrizing the phase boundary. $r_c$ is estimated from the $N$-independent crossing point of $[r]$ as shown in the inset. 
  Inset: Finite-size crossovers of mean level gap ratio $[r]$ as a function of energy density $\epsilon$ at fixed transverse field $\Gamma = 0.25$ at sizes $N=8,10,12,14$.
  (b) Finite-size crossovers of mean level gap ratio $[r]$ as a function of rescaled transverse field $\Gamma \sqrt{N} \log{N}$ at zero energy. The solid  vertical line gives the finite-size estimate of $\Gamma_c$ from Eq.~\eqref{eq:infinitedensity}. The horizontal dashed line at $[r] = 0.38$ ($0.53$) indicates the expected value for Poisson (GOE) level statistics.}
  \label{fig:fss-r}
\end{figure}

To map out the MBLD transition as a function of energy density and transverse field we first obtain the spectrum and many-body eigenstates of the Hamiltonian by numerical exact diagonalization. 
We use as diagnostics of the MBL phase both the spectral statistics and the dynamics of local observables\footnote{Due to the non-local nature of the model, only the $z$-component of spin behaves as a local observable.} and we find that they agree quantitatively in detecting two distinct phases: an ergodic and a MBL phase, sharply separated by a mobility edge (we see no evidence of a non-ergodic delocalized phase \cite{DeLuca:2014ch}). 
Focusing on the critical region, we find that the properties of the critical level statistics appear to vary continuously with energy density (see Figure \ref{fig:fss-r}). This raises the intriguing possibility of a continuous family of dynamical critical theories describing the MBLD transition in this model.
Analytical calculations are done in the forward-scattering approximation, by studying the statistical properties of the wave functions \cite{abou1973selfconsistent,nguyen1985tunnel,medina1992quantum,kardar1994lectures,altshuler1997quasiparticle, Mirlin:1997Tree, muller2013magnetoresistance,de2013support}. 
In particular we are able to see a transition from a MBL phase to a phase where resonances proliferate. This is the MBLD transition. The analytical and numerical results are in very good agreement and  can even estimate finite-size corrections for the critical transverse field based on the analytical calculations. 

The paper is organised as follows. We begin with presenting the evidence for the MBL phase and the MBLD transition based on the numerical results. 
Following which we introduce the forward scattering approximation and discuss its analytical consequences in relation to the numerics (additional details in the Supplemental Material). We conclude with a summary of the main results.

\paragraph{Thermodynamics---} 
The quantum random energy model (QREM) is defined by the following Hamiltonian on $N$ Ising spins
\begin{align}
\label{eq:ham}
	H =  E(\set{\hat{\sigma}_i^z}) - \Gamma \sum_{i=1}^N \hat{\sigma}_i^x,
\end{align}
where the first `classical REM' term is a random operator, diagonal in the $\sigma^z$ basis, while $\Gamma$ is a transverse field.
The $2^N$ diagonal energies $E(\set{\sigma^z})$ are independent and identically distributed Gaussian random variables with distribution 
\begin{align}
\label{eq:PE}
	P(E)=\frac{1}{\sqrt{\pi N}}e^{-\frac{E^2}{N}}.
\end{align}
Although typical energies $E$ from this distribution are of order $O(\sqrt{N})$, the full collection of $2^N$ independent energies produces an extensive spectrum. For instance, the ground state energy density of the classical REM is $E_0/N= \epsilon_0 = -\sqrt{\log(2)}$ with probability one (as $N\to\infty$). 
The thermal phase diagram at $\Gamma = 0$ follows immediately from the disorder averaged entropy function $s(\epsilon) = \log(2) - \epsilon^2$, as shown originally in \cite{Derrida:1980fk,Derrida:1981jd}.
The high temperature phase at $T>T_c = 1/2\sqrt{\log{2}}$ has the equilibrium properties of a classical paramagnet: no order exist and exponentially many states contribute to the partition function democratically. 
At $T_c$ a first order phase transition occurs into a `frozen' phase where an $O(1)$ number of states around the ground state dominate the partition function and the free energy density $f = \epsilon_0$ is a constant.

On increasing $\Gamma$, naive perturbation theory suggests that the energy density of all eigenstates is unchanged. 
Consequently, as is argued in \cite{Jorg:2008fj}, the free energy density is also unperturbed and the two classical phases extend to finite $\Gamma$ with a horizontal phase boundary. 
For sufficiently large $\Gamma$, however, the ground state is that of the transverse field term $\ket{QPM} = \ket{\rightarrow\cdots\rightarrow}$.  
Comparing the energy density $-\Gamma$ of this state to $\epsilon_0$ identifies a first order zero temperature quantum phase transition at $\Gamma_c = \sqrt{\log2}$ into the quantum paramagnet. 
A more detailed treatment \cite{Goldschmidt:1990p962} shows that this first order transition extends to infinite temperature, as does the quantum paramagnetic phase. 
The full thermodynamic phase diagram of the QREM is shown in Fig.~\ref{fig:pd}, the different thermodynamic phases are separated by dashed blue lines.

\paragraph{Quantum dynamics---} 

In this section we show how the QREM exhibits a MBLD transition consistent with the curve $\epsilon = \pm \Gamma$ in the micro canonical ensemble. Namely, the eigenstates with energy density $|\epsilon|>\Gamma$ are MBL, while if $|\epsilon|<\Gamma$ they appear to satisfy ETH. 

First, we provide a heuristic explanation of this behavior. In the large $\Gamma$ limit, where the spins are either aligned or anti-aligned with the transverse field, the spectrum separates into highly degenerate bands. 
The random energy term behaves as a perturbative random matrix in each of these bands giving rise to GOE level statistics. Thus, the quantum paramagnet is always thermal and the eigenstates satisfy the ETH; numerics shows that all the eigenstates within the energy window $\pm \Gamma$ are dominated by this extended behavior even on the classical side of the first order thermodynamic transition between the quantum and classical paramagnets. We will return to the analytic treatment which leads to the same conclusion.

Approaching from the delocalized side, and going from the micro canonical to the canonical ensemble allows us to define a critical temperature $T_{\text{MBL}} = 1/2\Gamma$, corresponding to the energy density $\epsilon=\pm\Gamma$ inside the classical paramagnetic phase.
That is, the system fails to thermalize throughout the low energy density regime (shaded red), and equilibrium statistical mechanics fails at temperatures well above the canonical spin glass transition $T_c$.

\begin{figure}[tb]
  \centering
    \includegraphics{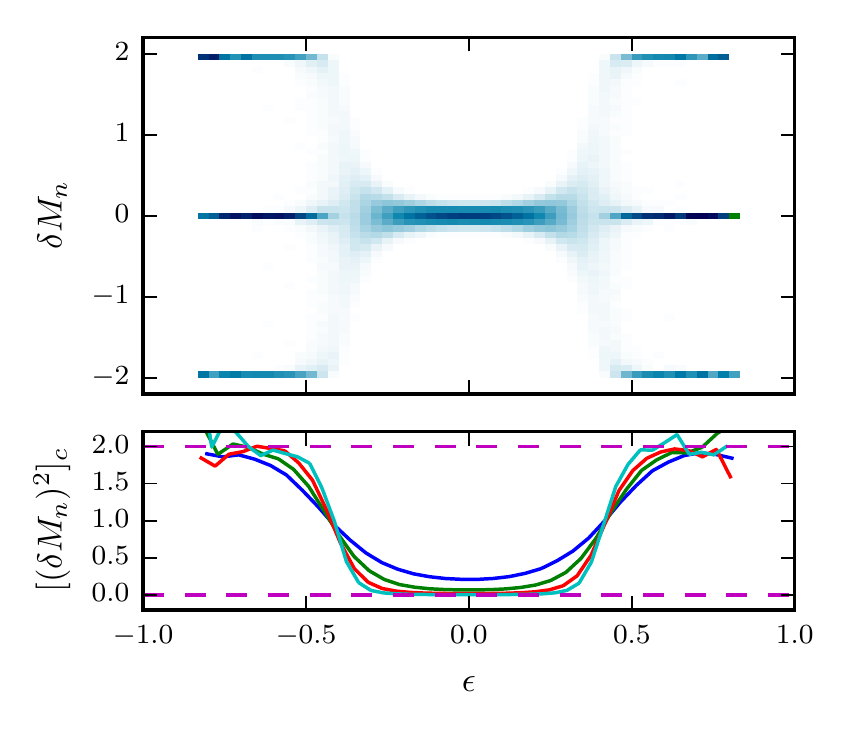}
  \caption{(color online) (a) Density plot of $P(\delta M_n)$ as a function of energy density $\epsilon$ at $\Gamma \approx 0.28$. Vertical line cuts are histograms across disorder and within narrow energy density windows. (b) Finite-size crossover of variance of $\delta M_n$ as a function of energy density.}
  \label{fig:spider}
\end{figure}

Numerically, we adduce several pieces of evidence in support of the conjecture that this curve corresponds to the MBLD transition. These include transitions in the many-body level statistics (Fig.~\ref{fig:fss-r}) and the presence of frozen local observables (Fig.~\ref{fig:spider}). 
All of these have been calculated within full exact diagonalization of systems with sizes $N=8,10,12,14$ with $N_s \approx 10^4-10^2$ samples per $\Gamma$ and per system size.
The statistics of gaps between many-body energies provide perhaps the simplest diagnostic.
We expect the delocalized phase to exhibit level repulsion following GOE random matrix theory while the MBL phase should exhibit Poisson statistics \cite{oganesyan2007localization}. 
These two regimes may be distinguished by using the level-spacing ratio $r_n^\alpha = \min\left\{\delta_n^\alpha,\delta_{n+1}^\alpha\right\} / \max\left\{\delta_n^\alpha,\delta_{n+1}^\alpha\right\}$, where $\delta_n^\alpha = E_n^\alpha - E_{n-1}^\alpha$ is the $n$-th gap between adjacent energy levels in a given sample $\alpha$. 
Taking the average over disorder and within narrow energy windows defines the mean level statistic $[r]$, which $\simeq 0.39$ for Poisson statistics or $\simeq 0.53$ for GOE statistics. 

The inset of Fig. 2(a) shows a typical example of the finite-size crossover of $[r]$ as a function of $\epsilon$ at $\Gamma=0.25$. 
The crossing point gives the critical energy density at which the eigenstates become delocalized, and it is the extracted values of these critical energies which are plotted in the phase diagram of Fig.~\ref{fig:pd}b.
At $\epsilon=0$ (infinite temperature), the $[r]$ curves for different $N$ as a function of $\Gamma$ do not cross (Fig.~\ref{fig:fss-r}b). 
Rather, the jump from Poisson to GOE level statistics becomes steeper and moves to smaller $\Gamma$ values as $N$ increases. 
This indicates that the infinite temperature eigenstates are delocalized for arbitrarily small $\Gamma$ in the thermodynamic limit but that the finite size flow of $\Gamma_{\text{MBL}}(N)$ is slow, in quantitative agreement with analytic estimates below. 
Finally, we note the continuous variation of the critical value of the crossing point $r_c$ along the critical boundary (Fig.~\ref{fig:fss-r}a). 
This variation suggests there may be a continuum of critical theories on the mobility edge.

The mean-field nature of the QREM complicates the study of local observables.
The random energy function $E(\set{\sigma^z})$ is highly non-local; indeed, typical spin configuration differing by $O(1)$ spin flips are $O(\sqrt{N})$ distant in energies. The transverse field term, on the other hand, is made up of a sum of local operators. 
Therefore the model retains a notion of locality as reflected in the commutators:
\begin{align}
	|[H, \sigma^z_i]| &= \Gamma \sim O(1) \nonumber\\
	|[H, \sigma^x_i]| &\sim O(N).
\end{align}
Thus, we expect on-site $\sigma^z$-magnetization to behave as a local observable while $\sigma^x$ does not. 
In an ergodic phase satisfying ETH, local observables evaluated in eigenstates of the Hamiltonian are smooth functions of the energy density $M_n = \bra{n}\sigma^z_0 \ket{n} \approx M(\epsilon_n)$ so that the difference in expectation values between two adjacent eigenstates,  $\delta M_n = M_n - M_{n-1} \approx M'(\epsilon_n) e^{-N s(\epsilon_n)}$, decays exponentially with $N$ where $s\left( \epsilon_n\right)$ is the entropy density of the states in the microcanonical energy shell. 
In the MBL regime, on the contrary, the magnetization varies by $O(1)$ between adjacent eigenstates. 

These features are reflected clearly in the `spider diagram' of Fig. 3(a), whose intensity shows the histogram of magnetization jumps $P(\delta M_n)$ as a function of energy density $\epsilon$ at size $N=14$, $\Gamma = 0.28$. 
Near zero energy density (infinite temperature), the body of the spider reflects the peak near $0$ of $P(\delta M_n)$ in the ergodic phase while the legs reflect the glassy freezing of $M_n \approx \pm 1$ in adjacent MBL eigenstates. 
The variance of the distribution of $\delta M_n$ shows finite-size scaling behavior which can also be used to estimate the critical energy density (Fig.~\ref{fig:spider}b).
In the ergodic phase $[(\delta M_n)^2]_c \rightarrow 0$ while in the localized phase it tends to 2 as shown in Fig. 3(b). 
The crossing, after finite size scaling is performed can be used to locate the MBLD transition.
The location of the MBLD transition detected by $[r]$ and $[(\delta M_n)^2]$ agree within error bars.

\paragraph{Perturbation theory and the structure of the wave functions--}
For small $\Gamma$, it is useful to think of the Hamiltonian \eqref{eq:ham} as defining a single-particle Anderson localization problem on the $N$-dimensional hypercube defined by the $\sigma^z$ basis states. 
In this picture, the $E(\{\sigma^z\})$ term is a random chemical potential on the vertices of the hypercube while the transverse field hops between adjacent vertices. The localization problem on the hypercube shares certain facets with that of a Bethe lattice with high branching number, but the hypercube possesses many short loops which are absent in the Bethe lattice.
In the MBL phase, the eigenstates remain close to an unperturbed $\sigma_z$ configuration (its \emph{localization center}), while the MBLD transition is signaled by the proliferation of resonances at large Hamming distance from the localization center \cite{altshuler1997quasiparticle, Mirlin:1997Tree}.

To leading order in $\Gamma$, the amplitude for a wavefunction concentrated on a spin configuration $a$ at $\Gamma=0$ to reach spin configuration $b$ at distance $n$-spin flips away is given by
\begin{align}
	\label{eq:forwardamplitude}
	\psi_b\simeq\Gamma^n\sum_{p\in \Pi_n}\prod_{i\in p}\frac{1}{E_a-E_i}	
\end{align}
where $p$ runs over the $n!$ shortest paths $\Pi_n$ from $a$ to $b$. 
These span a small hypercube of diameter $n$, which contains all the sites in between $a$ and $b$.
The forward scattering approximation \cite{abou1973selfconsistent,nguyen1985tunnel,medina1992quantum,kardar1994lectures,altshuler1997quasiparticle,muller2013magnetoresistance,de2013support}  consists in taking this leading order expression to define the amplitude at any given site $b$, thus neglecting higher order corrections from longer (loopy) paths. 

Consider the case of a finite temperature initial state in which $E_a = - \epsilon N$, $\epsilon > 0$. 
In this case, the $M = \sum_{j=1}^n \binom{N}{j}$ vertices within a distance $n$ of $a$ have energy in the range $\pm E^* = \sqrt{N}\sqrt{\ln M} \sim \sqrt{N\log N} \ll N \epsilon$ and thus the weights on all sites are typically of order $w_i = \frac{\Gamma}{N\epsilon} + O(N^{-2})$. 
Therefore these amplitudes sum coherently over the $n!$ paths leading to the given vertex $b$ giving $\psi_b \simeq n! \left(\frac{\Gamma}{N\epsilon}\right)^n$. 
This approximation neglects the small denominators which, for $n=O(N)$, start to appear, so we expect it to provide an underestimate of the probability of having a resonance $\psi_b\sim 1$. 
Nonetheless, we already find that for $n > n^* = N e \epsilon/\Gamma$, the wavefunction at site $b$ is $\sim 1$.
 
Requiring that the wavefunction be small throughout the hypercube ($n^* = N$), we find $\Gamma_{\text{MBL}} \le e \epsilon$. 
A more careful treatment of the probability of resonance at the $n+1$-th step given that the first $n$ steps are non-resonant \cite{altshuler1997quasiparticle} (see supporting material), gives a better estimate of the critical field. For
\begin{equation}
\label{eq:finitedensity}
\Gamma\leq\Gamma_{\text{MBL}} \simeq \epsilon+\sqrt{2}\epsilon^2+\frac{4}{3}\epsilon^3+...\ 
\end{equation}
the corresponding eigenstate will be many-body localized. Within this perturbative argument, the first resonance arises at distance $n^*=N(\sqrt{2}\epsilon-2\epsilon^2/3+...)$. 
Thus, as $\epsilon\to 0$ (infinite temperature) the first resonance approaches the initial site $a$ so that we need to treat the case of infinite temperature more carefully. 

Starting from an infinite temperature spin configuration $a$ ($\epsilon = 0$), the typical denominators are of order $\sqrt{N}$, and can become very small. The sum over paths from $a$ to $b$ will be dominated by only a small number of paths and constructive interference is not important. Bounding the probability of a rare resonances after $n \rightarrow N$ steps provides a finite-size estimate of the maximal strength of the transverse field $\Gamma_{\text{MBL}}$ for which the wavefunctions are localized (see Supplemental material). For large $N$, one finds 
\begin{equation}
\label{eq:infinitedensity}
\Gamma_{\text{MBL}}\simeq\sqrt{\pi}/2e\sqrt{N}\ln(e N)
\end{equation}
which tends to $0$ when $N\to\infty$. The scaling is in good agreement with the finite size flow of the numerics in Fig.\ref{fig:fss-r}b. 

Perturbation theory also suggests that the nature of the MBL eigenstates, including those at criticality, varies strongly with $\epsilon$. 
For $\epsilon>0$, the resonances occur every $n^*\sim N\sqrt{2} \epsilon $ hops. 
The critical states are therefore isotropic for large patches until a resonance is encountered. 
The overlap between neighbouring energy eigenstates are large and as an effect of this we expect $r_c$ to increase as we move away from the center of the spectrum, in qualitative agreement with Fig.~\ref{fig:fss-r}a. 
A systematic study of this quantity and the form of Chalker's scaling \cite{chalker1988scaling} at criticality is left for future work.
In the opposite limit, as $\epsilon \to 0$, close resonances proliferate, the wave functions are extremely irregular  and the critical statistics approach the Poisson value (similarly to the Bethe lattice Anderson model \cite{mirlin1991localization,de2013support}). Also this is seen in Fig.~\ref{fig:fss-r}a. A similar structure of MBL wave functions is observed in other systems \cite{de2013ergodicity,buccheri2011structure}. 

The statistical properties of the wavefunctions can also be studied using the replica method, which provides complementary understanding \cite{Mezard:1987aa,Mezard:2009aa,kardar1994lectures}. Leaving the details to the Supplementary Material, we find that the $\epsilon=0$, infinite temperature situation is in a one-step replica-symmetry breaking phase with a critical value of the field $\Gamma_{\text{MBL}}$ whose scaling is consistent with the direct analysis of Eq.~\eqref{eq:infinitedensity}.

\paragraph{Conclusions--} We have presented evidence, both numerical and analytical, for a MBLD transition to occur in the QREM independent from the equilibrium glass transition observed in the thermodynamics \cite{Goldschmidt:1990p962,Jorg:2008fj}. 
The QREM provides an analytically tractable mean-field type model for the MBLD transition.
Its local magnetization and level statistics behave in accordance with the expectations of MBL and ETH phenomenology.
The wavefunctions and level spacing statistics at criticality change their properties as the energy density changes along the MBLD phase boundary suggesting the existence of a continuous family of critical theories.

\emph{Acknowledgements--} We would like to thank M.\ Aizenman, B.\ Altshuler, V.\ Oganesyan, and S.\ Warzel for discussions and in particular A.\ Chandran, and R.\ Moessner for many useful comments on the first version of this manuscript. C.R.L.\ would like to acknowledge the hospitality of the Perimeter Institute where much of this work was undertaken. A.P.\ is supported by the Intelligence Advanced Research Projects Activity (IARPA), through the Army Research Office Grant No. W911NF-12-1-0354. A.S.\ is supported in part by NSF Grant No. PHY-1005429.

\bibliography{qremmbl}

\newpage
\section*{SUPPLEMENTAL MATERIAL}

We show how one can get the estimates (5) and (6) for the mobility edge by analyzing the perturbative wavefunctions in the forward approximation. 

At finite energy density $\epsilon=E_a/N$ resonances, i.e.\ values of the energy denominators $\delta_n=E_a-E_i$ particularly small, are quite rare. One has to go a distance of $O(N)$ to find such a resonance. Let us suppose the first resonance occurs at distance $n$. 
Prior to the resonance, the paths sum coherently with each contributing a typical value, so that we estimate
\begin{align}
\psi_{n-1} &\simeq (n-1)! \left(\frac{\Gamma}{\epsilon N}\right)^{n-1}
\end{align}
for all amplitudes at distance $n-1$. 
The are $n$ ways to reach a site at distance $n$ from sites at distance $n-1$. Hence,
\begin{eqnarray}
\psi_n&=&\frac{\Gamma}{\delta_n}n \psi_{n-1}
\end{eqnarray}
We have a resonance when $|\psi_n|>1$, namely if
\begin{equation}
|\delta_n|<\Gamma n |\psi_{n-1}|
\end{equation}
Therefore the (small) probability to have a resonance is
\begin{eqnarray}
p&=&\int_{\epsilon-\Gamma n |\psi_{n-1}|}^{\epsilon+\Gamma n |\psi_{n-1}|}d\epsilon \sqrt{\frac{N}{\pi}}e^{-N\epsilon^2}\nonumber\\
&\simeq& 2\Gamma n |\psi_{n-1}| \rho(\epsilon)
\label{eq:smallp}
\end{eqnarray}
with $\rho=\sqrt{\frac{N}{\pi}}e^{-\epsilon^2 N}$ comes from the distribution of levels.
Define $P_n$ as the probability that none of the $\binom{N}{n}$ points at level $n$ gives a resonance. In order to proceed we need to assume that these events are uncorrelated. This is an approximation which gives a lower bound to the probability $P_n$ and we will see how good this is compared to the numerical data. 
In this approximation,
\begin{equation}
P_n=(1-p)^{\binom{N}{n}}
\end{equation}
which, inserting (\ref{eq:smallp}) gives
\begin{eqnarray}
P_n&=&\left(1-2\Gamma n! \rho\left(\frac{\Gamma }{ \epsilon N}\right)^{n-1}\right)^{\binom{N}{n}}\nonumber\\
&\simeq&e^{-e^{Nf(x,\epsilon)}}
\end{eqnarray}
where $x\equiv n/N$ and,
\begin{equation}
f=N^{-1}\ln\left(2 \binom{N}{n} \Gamma n! \rho\left(\frac{\Gamma }{ \epsilon N}\right)^{n-1} \right). 
\end{equation}

We obtain to leading order in $1/N$,
\begin{equation}
\label{eq:fx}
f(x,\epsilon)=-(1-x)\ln(1-x)-\epsilon^2+x\ln\left(\frac{\Gamma}{e\epsilon}\right).
\end{equation}
As $N\to\infty$, if $f<0$ we have $P_n\to 1$, while for $f>0$ we have $P_n\to 0$. In order to see where the first resonance occurs we need to find the smallest $n$ such that $P_n=0$, so we have to find  $f^*$, the maximum of the function $f(x,\epsilon)$ over $x$ for any given $\epsilon$.
\begin{figure}[htbp]
\begin{center}
\includegraphics[width=0.45\columnwidth]{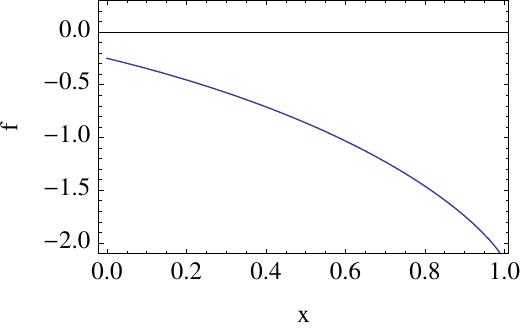}
\includegraphics[width=0.45\columnwidth]{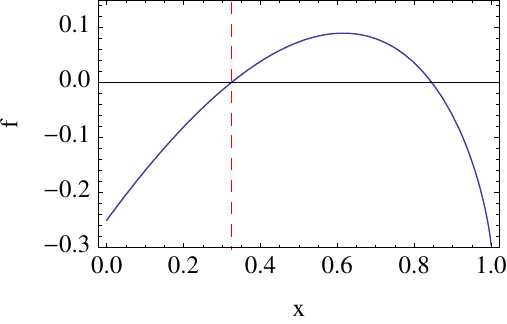}
\caption{(color online) The function $f$ in (\ref{eq:fx}) for $\epsilon=0.5$ and $\Gamma<\Gamma_c$ (left) and $\Gamma>\Gamma_c$ (right). In the right panel the red, dashed line is the position of $x^*=n^*/N$.}
\label{fig:fx}
\end{center}
\end{figure}
After some algebra we find 
\begin{equation}
f^*=\frac{\epsilon}{\Gamma}+\ln\left(\frac{\Gamma}{e\epsilon}\right)-\epsilon^2.
\end{equation}
Solving $f^*=0$ for $\Gamma$ we find an explicit form for $\Gamma_c(\epsilon)$ which can be expanded for small $\epsilon$ as 
\begin{equation}
\Gamma_c=\epsilon+\sqrt{2}\epsilon^2+\frac{4}{3}\epsilon^3+...\ .
\end{equation}
At constant $\Gamma$, varying $E_a$ therefore defines a many-body mobility edge. It is also instructive to see at which value of $n$ the maximum occurs, which gives the most probable position of the first resonance:
\begin{equation}
n^*=N\left(1-\frac{\epsilon}{\Gamma}\right)\simeq N(\sqrt{2}\epsilon-2\epsilon^2/3+\epsilon^3/(9\sqrt{2})...).
\end{equation}
We see from here that for any finite $\epsilon$ the position of the first resonance is at $O(N)$ away. As $\epsilon\to 0$ the first resonance comes quite close to the origin of the locator expansion. If we want to find the finite-$N$ corrections to $\Gamma_c$ at $\epsilon=0$, we need to consider this possibility more carefully. 

This leads to the discussion of the case at infinite temperature, i.e.\ where $E_a=0$. Let us
define the variable $y_i=-\ln(|E_i|/\sigma)$ for some $\sigma$ which we will fix shortly. $y_i\to\infty$ at a resonance $E_i=E_a=0$. 

We find
\begin{equation}
P(y_i)=\frac{2\sigma}{\sqrt{\pi N}}e^{-y_i-\frac{\sigma^2}{N}\exp(-2y_i)}.
\end{equation}
We choose now
\begin{equation}
\sigma=\frac{\sqrt{\pi N}}{2}
\end{equation}
so, since we are interested in rare fluctuations where $y_i\gg 1$, we have that
\begin{equation}
P(y_i)\simeq e^{-y_i} \quad\mbox{for } y_i\gtrsim 1.
\end{equation}
We need to study the distribution of the amplitudes 
\begin{equation}
A_p=\prod_{i=1}^n\frac{\Gamma}{0-E_i},
\end{equation}
over all the paths $p$ which go out to distance $n$. Consider all the $\mathcal{N}\equiv\prod_{i=0}^{n-1}(N-i)$ paths that go out to one of the $\binom{N}{n}$ points. They appear clustered in sums but since the distribution of their contributions is very large this does not matter: $O(1)$ of the paths will dominate both the sum to get to the point $b$ and the total probability of resonance at distance $n$. To control the latter, we will look for the probability that \emph{none} of these paths gives resonance. 
We already know that the first path to break this condition will be similar to the greedy path but performing the calculation will give an extra $\ln N$ correction, typical of Anderson localization problems on large connectivity graphs \cite{abou1973selfconsistent}.

Consider the log amplitude $A_p$ of a given path,
\begin{equation}
\ln |A_p|=n\ln (\Gamma/\sigma)+\sum_{i=1}^n y_i.
\end{equation}
We have a resonance if
\begin{equation}
|A_p|>1,
\end{equation}
namely if
\begin{equation}
\sum_{i=1}^n y_i>n\ln(\sigma/\Gamma)\equiv Y_c.
\end{equation}
Introducing $Y=\sum_{i=1}^n y_i$ one finds that it is distributed as
\begin{equation}
P(Y)=\frac{Y^{n-1}}{(n-1)!}e^{-Y},
\end{equation}
and so we have now all the ingredients to find the probability to have a resonance $|A_p|>1$ at distance $n$ (see also \cite{altshuler1997quasiparticle,abou1973selfconsistent,de2013support}).

Since $P(|A_p|>1)=P\left(Y>Y_c\right),$ where $Y_c=n\ln\left(\frac{\sigma}{\Gamma}\right)\gg 1$ we have 
\begin{eqnarray}
p\equiv P(Y>Y_c)&=&\int_{Y_c}^\infty dY\frac{Y^{n-1}}{(n-1)!}e^{-Y}\nonumber\\
&\simeq&\frac{Y_c^{n-1}}{(n-1)!}e^{-Y_c}+O(Y_c^{(n-2)})
\end{eqnarray}
doing the integral by parts. Using Stirling's approximation:
\begin{eqnarray}
p&\simeq& \frac{Y_c^ne^n}{n^n}e^{-Y_c}\nonumber\\
&=&\exp\left[-n\phi\left(\frac{\sigma}{\Gamma}\right)\right],
\end{eqnarray}
where $\phi(x)=\ln(x/(e\ln x))\geq 0$. 
Again, assuming that all $\mathcal{N}$ paths resonate independently (an underestimate), the probability that we do not have any resonant paths is 
\begin{equation}
(1-p)^{\mathcal{N}}\simeq e^{-\mathcal{N}p}.
\end{equation}
If $\mathcal{N}p\gg 1$ then the probability that no resonating path exists goes to zero. Defining $f=\ln(\mathcal{N}p)/n$ we have the condition 
\begin{eqnarray}
f&=&\frac{1}{n}\ln \mathcal{N}-\phi(\sigma/\Gamma)\nonumber\\
&\simeq&\ln N-\ln\left(\frac{\sigma}{e\Gamma\ln(\sigma/\Gamma)}\right)=0,
\end{eqnarray}
the condition for the transition gives
\begin{equation}
\frac{\sigma}{e\Gamma_c\ln(\sigma/\Gamma_c)}=N.
\end{equation}
The numerical solution of this equation for $N=8,...,14$ are reported in the text. We cannot solve this equation for $\Gamma_c$ exactly but in the large $N$ limit, the solution is
\begin{equation}
\Gamma_c\simeq \frac{\sigma}{eN\ln eN}=\frac{\sqrt{\pi}}{2eN^{1/2}\ln (eN)}+O\left(\frac{1}{N^{1/2}\ln^2 N}\right),
\end{equation}
as quoted in the main text.

\paragraph{Replica treatment--} 
The statistical properties of the wavefunctions can also be studied using the replica method, which provides complementary but non-rigorous understanding \cite{Mezard:1987aa,Mezard:2009aa}. 
In this approach, we view the amplitude $\psi_b$ as the partition sum of a directed random polymer (the path) living on the \emph{hypercube} with the long-tailed random weights $w_i = \Gamma/(E_a - E_i)$. 
Notice that these weights do not have any finite moments so we expect the directed random polymer to condense onto a small number of large weight paths \cite{kardar1994lectures}.
We focus on the most interesting case of infinite temperature states, where the replica approach is most useful as it naturally regulates the divergence of the weights.

The typical value of the forward scattering wavefunction $f = \overline{\ln |\psi|}$ admits a straightforward replica treatment exploiting the usual relationship $\overline{\ln |\psi|} = \Re\lim_{m\to0} \frac{\overline{\psi^m} - 1}{m}$.
In the 1RSB ansatz, the dominant configurations contributing to $\overline{\psi^m}$ consist of $m/x$ tightly bound groups of $x$ paths each. This gives rise to the 1RSB free energy: 
\begin{align}
  f(x) = \frac{n}{x}\left(\log n - 1 + \log \overline{w_i^x}\right)
\end{align}
where $x \in [0,1]$ is the Parisi parameter and $w_i = \Gamma/(E_a - E_i)$ is the weight on site $i$.
Minimizing over $x$, we find that the saddle point of the replicated free energy arises at $x^* = 1- \frac{1}{\log \sqrt{2/\pi} n} + \cdots$ as $n\to\infty$, indicating condensation of the paths.

Solving for the resonance condition $\Re f = 0$ at $n=N$, we find the estimate
\begin{equation} 
\Gamma_c = \frac{\sqrt{\pi}}{2\sqrt{N}\log \sqrt{2/\pi}N} + \cdots
\end{equation}
for the critical value of the transverse field. 
We note that this estimate is larger by a factor of $e$ than the estimate from the direct probabilistic analysis above, but it has the same scaling with $N$. 
This is natural as the resonance condition used here is that the amplitude at the far side of the hypercube should diverge as opposed to a small (but entropic) collection of atypical resonances appearing somewhere in the cube, as estimated above. 

\end{document}